\documentclass[aps,prl,twocolumn,showpacs,amsmath,amssymb,superscriptaddress,longbibliography]{revtex4-1}

\usepackage[english]{babel}
\usepackage{amsmath,amsfonts,bm,graphicx,verbatim,mathrsfs,units,color}
\usepackage[colorlinks=true,citecolor=blue]{hyperref}
\usepackage{pdfpages}

\newcommand{\ket}[1]{\left| #1 \right \rangle}

\newcommand{\revision}[1]{{#1}}

\begin{document}
\title{Dynamical Scheme for Interferometric Measurements of Full-Counting Statistics}
\date{\today}

\author{David Dasenbrook}
\affiliation{D\'epartement de Physique Th\'eorique, Universit\'e de Gen\`eve, 1211
  Gen\`eve, Switzerland}
\author{Christian Flindt}
\affiliation{Department of Applied Physics, Aalto University,
  00076 Aalto, Finland}

\begin{abstract}

We propose a dynamical scheme for measuring the full counting statistics in a mesoscopic conductor
using an electronic Mach-Zehnder interferometer. The conductor couples capacitively to one arm of
the interferometer and causes a phase shift which is proportional to the number of transferred
charges. Importantly, the full counting statistics can be obtained from average current measurements
at the outputs of the interferometer. The counting field can be controlled by varying the time delay
between two separate voltage signals applied to the conductor and the interferometer, respectively.
As a specific application we consider measuring the entanglement entropy generated by partitioning
electrons on a quantum point contact. Our scheme is robust against moderate environmental dephasing
and may be realized thanks to recent advances in giga-hertz quantum electronics.

\end{abstract}

\maketitle

\paragraph{Introduction.---} Full counting statistics (FCS) is a central concept in mesoscopic physics \cite{levitov96,blanter00,nazarov03}. The distribution of charge transfers contains information about the elementary conduction processes \cite{vanevic07,vanevic08,hassler08,abanov08,abanov09}. Full counting statistics has found widespread use in theories of quantum electronic circuits, for instance in proposals for detecting entanglement~\cite{beenakker04,dilorenzo05}, revealing interactions~\cite{kambly11,stegmann15}, understanding quasi-probabilities~\cite{belzig01,bednorz10,clerk11,hofer16}, or observing Majorana modes~\cite{soller14,li15,gnezdilov15,liu15,strubi2015}. Intimate connections to fluctuation relations at the nano-scale \cite{tobiska05,forster08,esposito09,utsumi09,nagaev10,utsumi10,kung12,saira12} and to entanglement entropy in fermionic many-body systems \cite{klich09,song11,song12,petrescu14,thomas15} have also been discovered.

Despite these promising applications, experiments remain scarce. Measurements of FCS are demanding as they require accurate detection of rare events in the tails of the distributions. For quantum-dot systems, progress has been made using real-time charge detectors~\cite{fujisawa06,gustavsson06,flindt09,gustavsson09,ubbelohde12,maisi14}. By contrast, for phase-coherent transport in mesoscopic conductors, only the first few cumulants of the current have been measured \cite{reulet03,bomze05,timofeev07,gershon07,gabelli09}. To measure the FCS, it has been suggested to use a spin to sense the magnetic field generated by the electrical current in a mesoscopic conductor \cite{levitov96,lesovik06,lebedev16}. However, being experimentally challenging, this proposal has not yet come to fruition.

Now, progress in giga-hertz quantum electronics is changing these perspectives \cite{bocquillon14}. Coherent electrons can be emitted on demand from quantum capacitors~\cite{gabelli06,feve07} and clean single-particle excitations can be generated using Lorentzian voltage pulses \cite{dubois13,jullien14}. In parallel with these developments, electronic interferometers have emerged as powerful detectors of weak signals~\cite{henny99,oliver99}. Mach-Zehnder interferometers can be realized using quantum Hall edge states with quantum point contacts (QPCs) acting as electronic beam splitters \cite{ji03,neder07nat,neder07,neder07prl,roulleau08,litvin10,helzel15}. When combined, these building blocks may form the basis for the next generation of quantum electronic circuits, including future measurements of FCS.

\begin{figure}
  \centering
  \includegraphics[width=0.85\columnwidth]{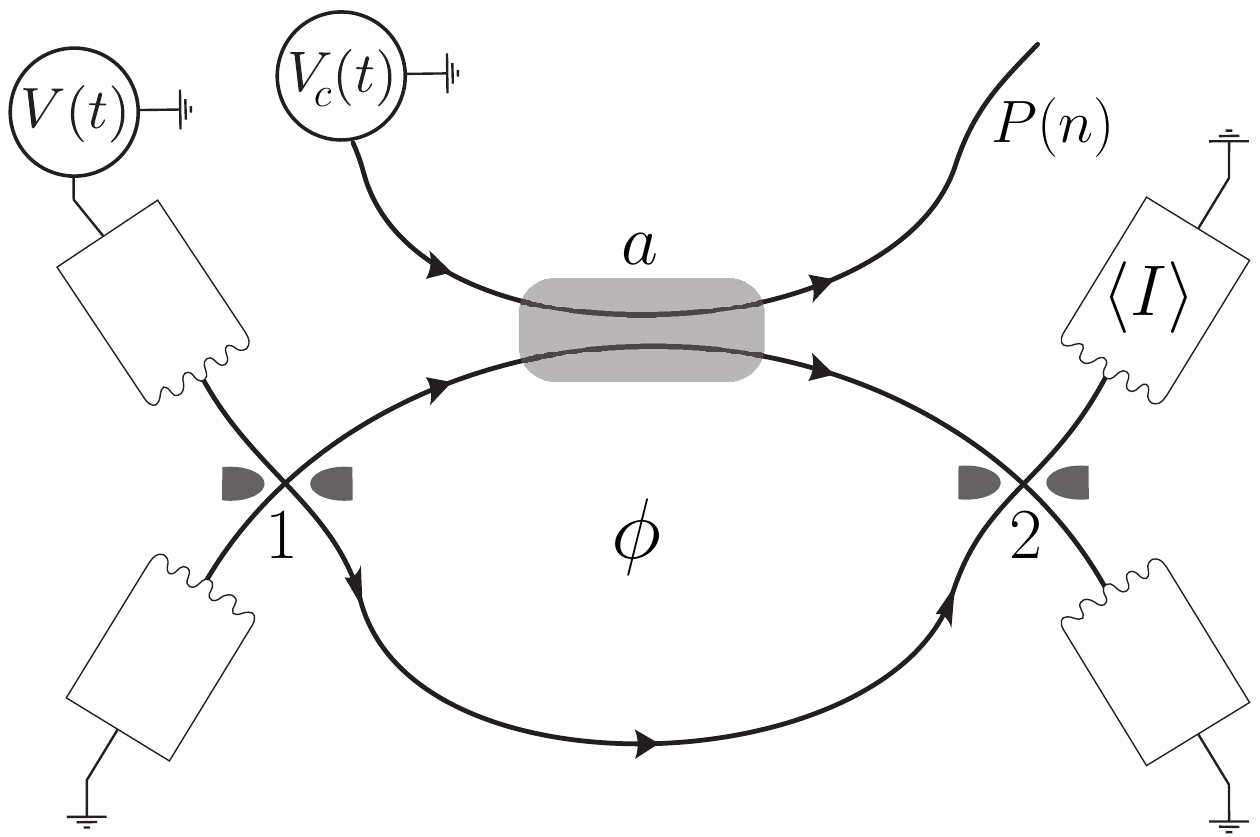}
  \caption{Interferometric measurements of FCS. Single electrons are injected into a Mach-Zehnder interferometer enclosing the magnetic flux $\phi$. \revision{Separate voltage signals are applied to the interferometer $V(t)$ and a nearby conductor $V_c(t)=V(t-\tau)$ with a time delay $\tau$.} The average current \revision{$\langle I\rangle $} measured at an output is sensitive to a phase shift caused by the capacitive coupling to the conductor \revision{in the interaction region of length $a$}. The phase shift is proportional to the number of transferred electrons in the conductor. By varying the magnetic flux and the time delay \revision{$\tau$}, the FCS of the conductor can be obtained from average current measurements only.}
  \label{fig:setup}
\end{figure}

Motivated by these experimental advances we develop in this Letter a dynamical scheme for measuring the FCS in mesoscopic conductors. The detector consists of an electronic Mach-Zehnder interferometer driven by periodic voltage pulses~\cite{hofer14,gaury14}. One arm of the interferometer is capacitively coupled to a nearby conductor that causes a phase shift which is proportional to the number of transferred charges, see Fig.~\ref{fig:setup}. As we will see, the FCS of the conductor can be inferred from current measurements at the outputs of the interferometer. Setups of this type, with static voltages, have been considered both in experiment \cite{neder07} and theory \cite{neder07njp,levkivskyi09}. However, so far the conductor has been operated as a which-path detector for the interferometer \cite{neder07prl,dressel12,weisz14}. Here, by contrast, we exchange the roles and instead use the interferometer as a \emph{detector} of the FCS in the conductor. \revision{Specifically, using a microscopic theory we derive an effective coupling for the interaction between the interferometer and the conductor which functions as the counting field of the FCS. The coupling can be controlled by varying the delay between separate voltage signals applied to the interferometer and the conductor. Importantly, the FCS is independent of the width of the applied pulses, and our proposal thus encompasses measurements of the FCS with very wide pulses corresponding to a constant voltage.}

\paragraph{Mach-Zehnder interferometer.---} The interferometer is implemented with edge states of a two-dimensional electron gas in the integer quantum Hall regime
\cite{ji03,neder07nat,neder07,neder07prl,roulleau08,litvin10,helzel15}. Incoming electrons are coherently split at the first QPC and recombined at the second. Single electrons are emitted into the interferometer by applying periodic voltage pulses to one of the inputs~\cite{hofer14,gaury14,lebedev05,hassler07,keeling06}. The pulses are sufficiently separated so that only one electron at a time traverses the interferometer~\footnote{In an alternative implementation, one may consider the injection of charges from a mesoscopic capacitor}. The electronic state inside the interferometer is a coherent superposition of the electron being in the
upper ($\ket{u}$) or lower ($\ket{l}$) arm~\footnote{This state resembles the spin in the proposal
  of Ref.~\cite{levitov96}},
\begin{equation}
  \label{eq:flyingqubitstate}
  \ket{\Psi} = t_1 \ket{l} + e^{i\phi} r_1 \ket{u}.
\end{equation}
Here, $t_1$ and $r_1$ are the transmission and reflection amplitudes of the first QPC and $\phi = 2 \pi \Phi / \Phi_0$ is the ratio of the magnetic flux $\Phi$ enclosed by the arms over the flux quantum $\Phi_0$. For electrons injected into the interferometer with period $\mathcal{T}$, the current in the upper output reads
\begin{equation}
\langle \hat{I} \rangle = (e/\mathcal{T}) |t_1 t_2 + r_1 r_2 e^{i\phi}|^2,
\end{equation}
where $t_2$ and $r_2$ are the transmission and reflection amplitudes of the second QPC. Equation (\ref{eq:flyingqubitstate}) describes a pure state. More generally, for instance due to a finite temperature or external noise causing fluctuations of $\phi$, the electron must be described by a density matrix $\hat{\rho}$. Importantly, a measurement of the average current yields an ensemble average over the phase $\phi$ \cite{samuelsson06}.

\paragraph{Basic principle.---} The Mach-Zehnder interferometer is coupled to a nearby conductor whose current fluctuations we wish to measure. The electrical fluctuations are described by the moment generating function (MGF)
\begin{equation}
  \label{eq:fcsmgf}
  \chi(\lambda) = \sum_n P(n) e^{i n \lambda}=\langle e^{i n \lambda}\rangle.
\end{equation}
The average is defined with respect to the probability $P(n)$ of $n$ charges being transmitted through the conductor and  $\lambda$ is the counting field. The conductor is driven by periodic pulses such that the MGF after many periods ($N \gg 1$) factorizes as $\chi(\lambda) = [\chi_\text{ext}(\lambda)]^N$, where $\chi_\text{ext}(\lambda)$ characterizes the \emph{extensive} FCS per period~\cite{ivanov97}. We focus on the measurement of $\chi_\text{ext}(\lambda)$ and omit the subscript ``ext'' in the following.

The conductor is coupled to the upper arm of the interferometer. Such a setup has been experimentally realized~\cite{neder07,neder07prl,roulleau08}, albeit with statically biased contacts. By contrast, here we drive both the conductor and the interferometer with periodic voltage pulses. The frequency of the two pulse sequences is the same, but we allow for a time delay $\tau $ between them. With this setup, an electron in the upper arm picks up the additional phase $\delta \phi = n \lambda$ due to $n$ electrons passing by in the conductor per period. \revision{This connection is derived in a detailed analysis below, where we use a microscopic theory for the interaction between the interferometer and the conductor to show that the effective dimensionless coupling $\lambda$ indeed can be identified with the counting field.} At zero temperature, the density matrix of the interferometer reads
\begin{equation}
  \label{eq:reduceddm}
  \hat{\rho} =  \begin{pmatrix}
    |t_1|^2 & t_1 r_1^* e^{i \phi} \chi(\lambda) \\
    t_1^* r_1e^{-i \phi} [\chi(\lambda)]^* & |r_1|^2
  \end{pmatrix},
\end{equation}
having used $\langle e^{i\delta \phi}\rangle= \langle e^{in\lambda}\rangle=\chi(\lambda)$. The off-diagonal element of $\hat{\rho}$ contain the MGF of the
conductor. \revision{Additional dephasing due to other noise sources can be included in the off-diagonal elements as we discuss below.} Equation (\ref{eq:reduceddm}) generalizes (\ref{eq:flyingqubitstate}) to non-pure states. \revision{It corresponds to an average over many periods and, as such, does not have any particular periodicity or time-dependence.}

The MGF can now be extracted from the current in the upper output. The current $\langle \hat{I} \rangle = \operatorname{tr} [ \hat{\rho} \hat{I}]$ reads
\begin{equation}
  \label{eq:current1}
  \langle \hat{I} \rangle= (e/\mathcal{T})(T_1T_2+R_1R_2+2\operatorname{Re}\{t_1^*t_2^*r_1r_2 e^{i\phi}\chi(\lambda)\} ),
\end{equation}
where $T_j  = |t_j|^2$ and $R_j=|r_j|^2$ are the transmissions and reflections of the two QPCs ($j=1,2$). At half transmission, we get $\langle \hat{I} \rangle= (e/\mathcal{T})(1+\operatorname{Re}\{ e^{i\phi}\chi(\lambda)\} )/2$. Moreover, by changing the magnetic flux, we find $\langle \hat{I} \rangle_{\phi=0} = (e/\mathcal{T})(1+\operatorname{Re}\{\chi(\lambda)\} )/2$ and $\langle \hat{I}_1 \rangle_{\phi=3\pi/2} = (e/\mathcal{T})(1+\operatorname{Im}\{\chi(\lambda)\} )/2$. These expressions lead us to the MGF
\begin{equation}
  \label{eq:mgfmeasurement}
  \chi(\lambda) = \frac{2 \mathcal{T}}{e} \left[ \left(\langle \hat{I}  \rangle_{\phi=0}-\frac{e}{2 \mathcal{T}}\right) + i \left( \langle \hat{I} \rangle_{\phi=3\pi/2} -\frac{e}{2 \mathcal{T}}\right) \right].
\end{equation}
Remarkably, the MGF can be obtained from average current measurements. This is the first central result of our work. As we go on to show, the counting field $\lambda$ can be controlled by varying the time delay $\tau $ between the pulse sequences. We can then perform a full tomography of~$\chi(\lambda)$ and thereby evaluate the FCS of charge transfer.

\begin{figure*}
  \centering
  \includegraphics[width=0.95\columnwidth]{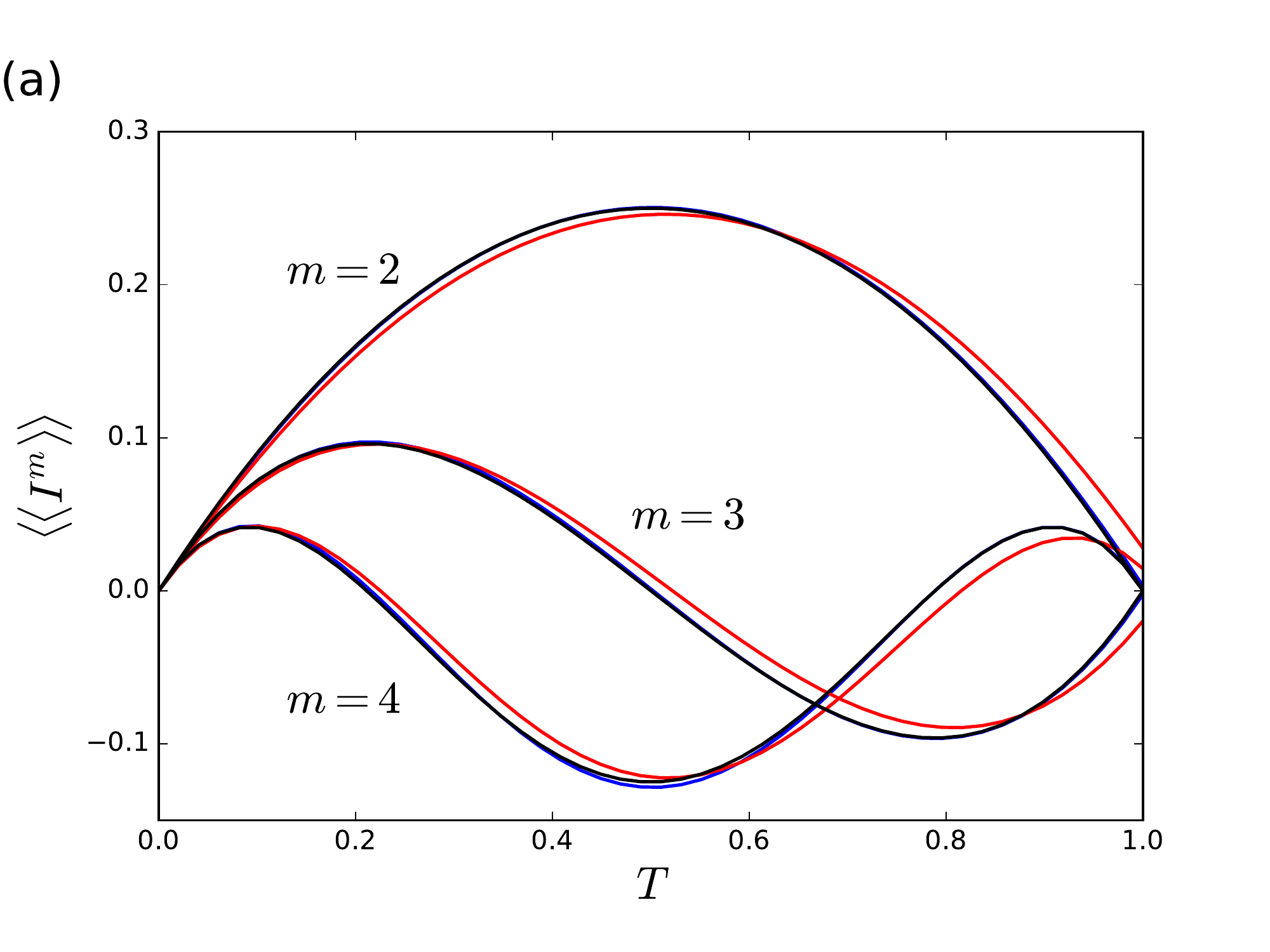}
  \includegraphics[width=0.95\columnwidth]{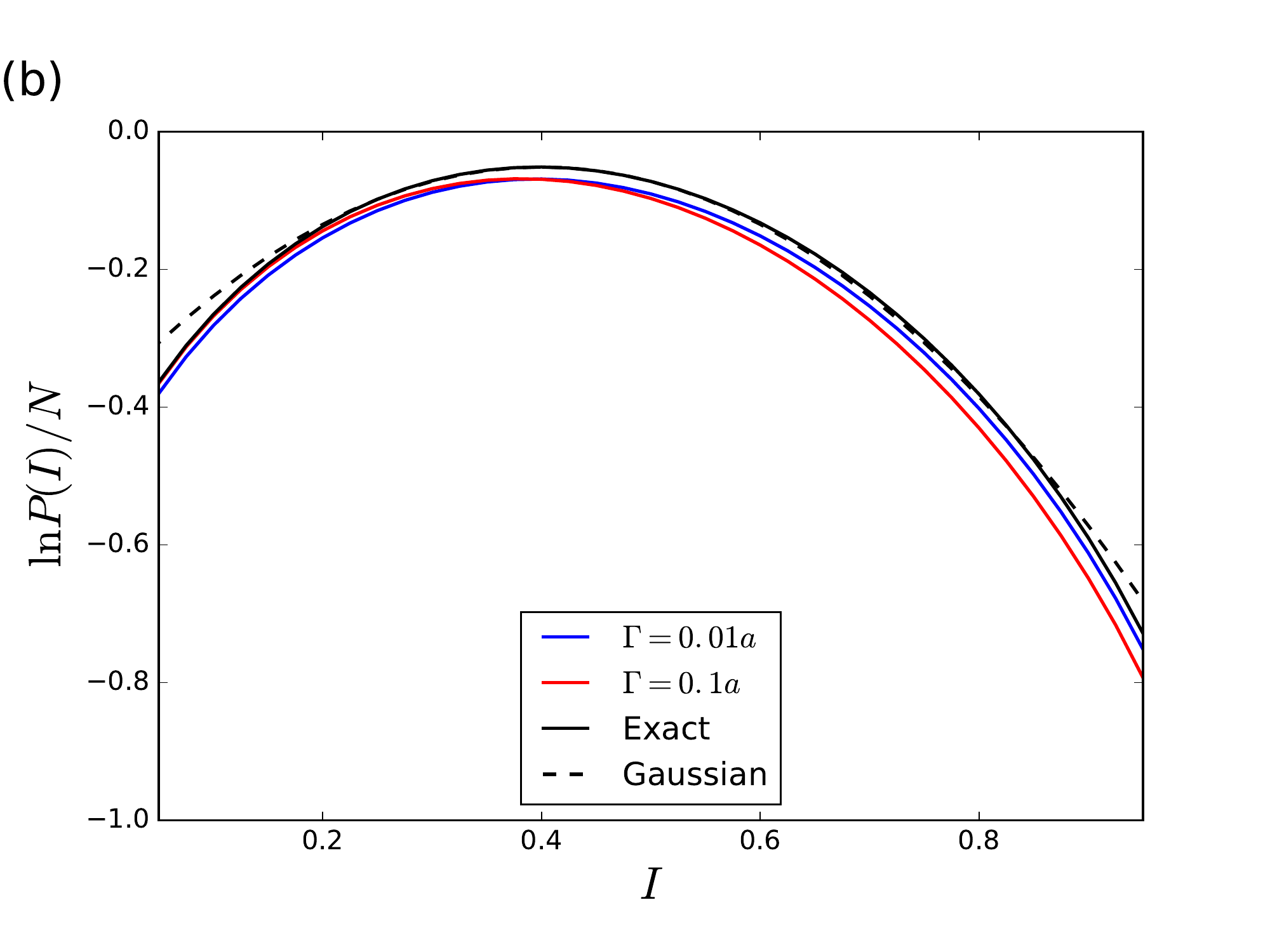}
  \caption{Interferometric measurement of the full counting statistics in a QPC. (a) Cumulants of the current as functions of the QPC transmission $T$. We show results for different pulse widths $\Gamma$ in terms of the length $a$ of the interaction region. The exact results for a binomial process are $\langle\!\langle I^2 \rangle\!\rangle = T(1-T)$, $\langle\!\langle I^3 \rangle\!\rangle=T(1-T)(1-2T)$, $ \langle\!\langle I^4 \rangle\!\rangle= T(1 - T)(1-6T + 6T^2)$, having set $e=1$ and $\mathcal{T}=1$ here and in the figure. (b) Full distribution of the current $I=en/N$ with $T=0.4$ and $N=40$. For a large number of periods, $N\gg 1$, the distribution takes on the large-deviation form $\ln[P(I)]/N=\mathcal{G}(I)$ with the rate function $\mathcal{G}(I)$ being independent of $N$. For a binomial process we find $\ln[P(I)]/N=\ln[(1-T)/(1-I)]+I(\ln[T/(1-T)]-\ln[I/(1-I)])+\mathcal{O}(N^{-1})$.}
  \label{fig:qpc}
\end{figure*}

\paragraph{Detailed analysis.---} We now embark on a detailed analysis of the coupling between the interferometer and the conductor. The interaction between the two edge states is described by the Hamiltonian~\cite{vyshnevyy13}
\begin{equation}
  \label{eq:hint}
  \hat{H}_{\lambda_0} =  \lambda_0\frac{\hbar v_F}{2a} \hat{N}_C\hat{N}_I.
\end{equation}
Here, $\lambda_0$ is a dimensionless coupling, $v_F$ is the Fermi velocity, and $a$ is a characteristic length scale over which electrons in the two edge states interact. The operators $\hat{N}_C =\int \mathrm{d}x \kappa_C(x)\!\!:\!\hat{\Psi}_C^\dagger(x) \hat{\Psi}_C(x)\!:$ and $\hat{N}_I =\int \mathrm{d}y \kappa_I(y)\!\! :\!\hat{\Psi}_I^\dagger(y) \hat{\Psi}_I(y)\!:$  count the number of excess electrons in the interacting regions of the conductor and the interferometer, weighted by the coordinate kernels $\kappa_{C}(x)$ and $\kappa_{I}(y)$. Normal-ordering with respect to the Fermi sea is denoted as $:\dots:$, and $\hat{\Psi}_{C}(x)$ and $\hat{\Psi}_{I}(y)$ are field operators for electrons in the conductor and in the interferometer.

The MGF in the off-diagonal element of the density matrix in Eq.~(\ref{eq:reduceddm}) can now be expressed as
\cite{levitov96,belzig01}
\begin{equation}
  \label{eq:offdiagmgf}
  \chi(\lambda_0) = \left \langle \operatorname{tr} \left(\widetilde{T} \left\{e^{-i \int_{t_0}^t \mathrm{d}t'\hat{H}_{\lambda_0}(t')
  }\right\} T\left\{ e^{i \int_{t_0}^t\mathrm{d}t' \hat{H}_{-\lambda_0}(t') }\right\} \hat{\rho}\right)   \right \rangle,
\end{equation}
having set $\hbar=1$ and $\hat{H}_{\lambda_0}(t)$ is in the Heisenberg representation governed by the full Hamiltonian $\hat{H}=\hat{H}_0 +\hat{H}_{\lambda_0}$ with $\hat{H}_0$ describing the uncoupled systems.  The initial density matrix of the electron in the interferometer is denoted as $\hat{\rho}=\hat{\rho}(t_0)$ and the trace is taken over the spatial coordinates. The average is defined with respect to the electrons in the conductor. Time and anti-time ordering are denoted as $T$ and $\widetilde{T}$, respectively.

The considerations above are general. To make further progress, we take for the kernels the specific form~\cite{vyshnevyy13}
\begin{equation}
  \label{eq:specifickappa}
  \kappa_C(x) = \kappa_I(x) = e^{-|x|/a}.
\end{equation}
If $a$ is much smaller than the length of the interferometer, the current measured at the output is
determined by the limit $t \to \infty$ in Eq.~\eqref{eq:offdiagmgf}. With a linear dispersion
relation for electrons close to the Fermi level and a pure initial state of the electron in the
interferometer, we find~\footnote{Details of this calculation can be found in the Supplemental
  Material at \ldots}
\begin{equation}
  \label{eq:offdiagmgf2}
  \chi(\lambda_0) = \left \langle \int \mathrm{d} y e^{i\lambda_0 \int \mathrm{d} x :\hat{\Psi}_C^\dagger(x)
  \hat{\Psi}_C(x): \Phi(x,y)} |f(y)|^2 \right \rangle,
\end{equation}
where $f(y)$ is the wave function of the electron injected into the interferometer and the function
\begin{equation}
  \label{eq:Phixy}
  \Phi(x,y) = e^{- \frac{|x-y|}{a}} \left( 1 + \frac{|x-y|}{a} \right)
\end{equation}
follows from the definition of the coordinate kernels.

We first consider the injection of electron wave packets with small widths compared to $a$, so that we can approximate $|f(y)|^2 \simeq \delta(y)$ and ${:\Psi_C^\dagger(x) \Psi_C(x):} \simeq \delta(x+v_F \tau) \hat{n}(x)$, where $\hat{n}(x)$ is the number operator for excess electrons in the conductor at position $x$ and $\tau$ is the time delay between the injection of electrons into the conductor and the interferometer. Equation~(\ref{eq:offdiagmgf2}) then yields
\begin{equation}
  \label{eq:vanishingwidth}
  \chi(\lambda) = \left \langle e^{i \hat{n}\lambda(\tau) } \right \rangle,
\end{equation}
with
\begin{equation}
  \label{eq:lambdatau}
  \lambda(\tau) = \lambda_0  e^{-v_F \tau/a} \left( 1 + \frac{v_F \tau}{a}
  \right).
\end{equation}
Equation (\ref{eq:lambdatau}) is the second important result of our work. It shows that the effective counting field $\lambda$ can be controlled by changing the time delay $\tau$. Negative values of the counting field can be realized by injecting hole-like excitations into the interferometer~\footnote{\revision{One can in principle use any range of counting fields of length $2\pi$, for instance $[0,2\pi]$ or $[-\pi,\pi]$. However, the maximal value of the counting field is given by the dimensionless coupling $\lambda_0$ in Eq.~(\ref{eq:lambdatau}). Thus, if this coupling is smaller than $2\pi$, one needs to use the interval $[-\pi,\pi]$}}. The specific functional form of Eq.~(\ref{eq:lambdatau}) is determined by the coordinate kernels in Eq.~(\ref{eq:specifickappa}) and, in reality, the dependence on $\tau$ may be different. Experimentally, one may then obtain $\lambda(\tau)$ using a conductor with a known FCS, e.~g.~a fully open QPC, for calibration.

In general, the wave functions have a finite width. Evaluating Eq.~\eqref{eq:offdiagmgf2} with the same wave functions $f(x)$ in the conductor and the interferometer, we find
\begin{equation}
  \label{eq:mgffinitedetector}
  \chi_\text{meas}(\lambda) = \int \mathrm{d} y |f(y)|^2 \chi(\widetilde{\lambda}(y,\lambda))
\end{equation}
with $\widetilde{\lambda}(y, \lambda) = \lambda \int \mathrm{d} x \Phi(x,y) |f(x)|^2$ and $\lambda$ given by Eq.~(\ref{eq:lambdatau}). Thus, for finite widths a measurement yields an average of MGFs for different effective couplings. However, if the pulses applied to the interferometer are sharper than the length of the interaction regions, we can incorporate the finite width of the electrons in the conductor into a rescaling of the effective counting field $\lambda$, which again can be obtained by proper calibration.

\paragraph{Driven quantum point contact.---} To illustrate our measurement scheme, we consider a QPC driven by Lorentzian voltage pulses of unit
charge as realized in recent experiments~\cite{dubois13,jullien14}. The QPC transmits electrons with probability $T$, and the exact MGF reads $\chi(\lambda) = 1 + T (e^{i \lambda}-1)$. The measured MGF is given by Eq.~\eqref{eq:mgffinitedetector} with a Lorentzian wave packet $|f(y)|^2 = 2 \Gamma / (y^2+\Gamma^2)$ of width $\Gamma$. We now obtain the cumulants of the current as
\begin{equation}
\langle\!\langle I^m\rangle\!\rangle = \frac{e^m}{\mathcal{T}} \partial_{i\lambda}^m \ln\{\chi_\text{meas}(\lambda)\}|_{\lambda\rightarrow 0},
\end{equation}
where $\lambda$ is the rescaled counting field. In Fig.~\ref{fig:qpc}a we show results for the cumulants as functions of the QPC transmission. For narrow wave packets, we find good agreement with analytic results for a binomial process.

Next, we turn to the full distribution of transferred charge after $N$ periods, given by the inversion formula
\begin{equation}
P(n)=\frac{1}{2\pi}\int_{-\pi}^{\pi}d\lambda e^{ N[\ln\{\chi_\text{meas}(\lambda)\}-i\lambda n/N]}.
\label{eq:fulldist}
\end{equation}
For a large number of periods, $N\gg1$, the distribution of the current $I=e n/(N\mathcal{T})$ takes on the large-deviation form $P(I)\simeq e^{\mathcal{G}(I)N}$ following from a saddle-point approximation of the integral in Eq.~(\ref{eq:fulldist}). Here, the rate function $\mathcal{G}(I)$ describes the exponentially rare current fluctuations, beyond what is captured by the central-limit theorem. In Fig.~\ref{fig:qpc}b, we again find good agreement with the analytic result for a binomial distribution.

\paragraph{Dephasing.---}
Our scheme is based on the reduced visibility in the Mach-Zehnder interferometer due to the dephasing induced by electrons in the conductor. In realistic systems, however, the visibility will already be reduced due to other dephasing mechanisms such as finite temperatures, the coupling to bulk electrons, co-propagating edge states, or electrons in the Fermi sea~\cite{levkivskyi08}. These effects are encoded in an additional fluctuating phase $\delta \theta$ \cite{ji03,roulleau09,bieri09}. \revision{It is reasonable to assume that the dephasing due to bulk phonons for instance is statistically independent from the dephasing due to the electrons in the conductor.} For a Gaussian distribution of width $\sigma$, the measured MGF then simply gets rescaled as $\chi_\text{meas}(\lambda)\rightarrow e^{-\sigma^2} \chi_\text{meas}(\lambda)$, and the width can be determined from a visibility measurement without electrons injected into the conductor.
For non-Gaussian fluctuations \cite{neder07njp,helzel15}, the total measured MGF takes the form $\chi_\text{meas}(\lambda) \rightarrow \chi_\text{meas}(\lambda) \chi_\text{env}(\lambda_0)$ for some fixed coupling $\lambda_0$ to the environment, such that the environmental contribution $\chi_\text{env}(\lambda_0)$ again can be factored out.

\begin{figure}
  \centering
  \includegraphics[width=0.8\columnwidth]{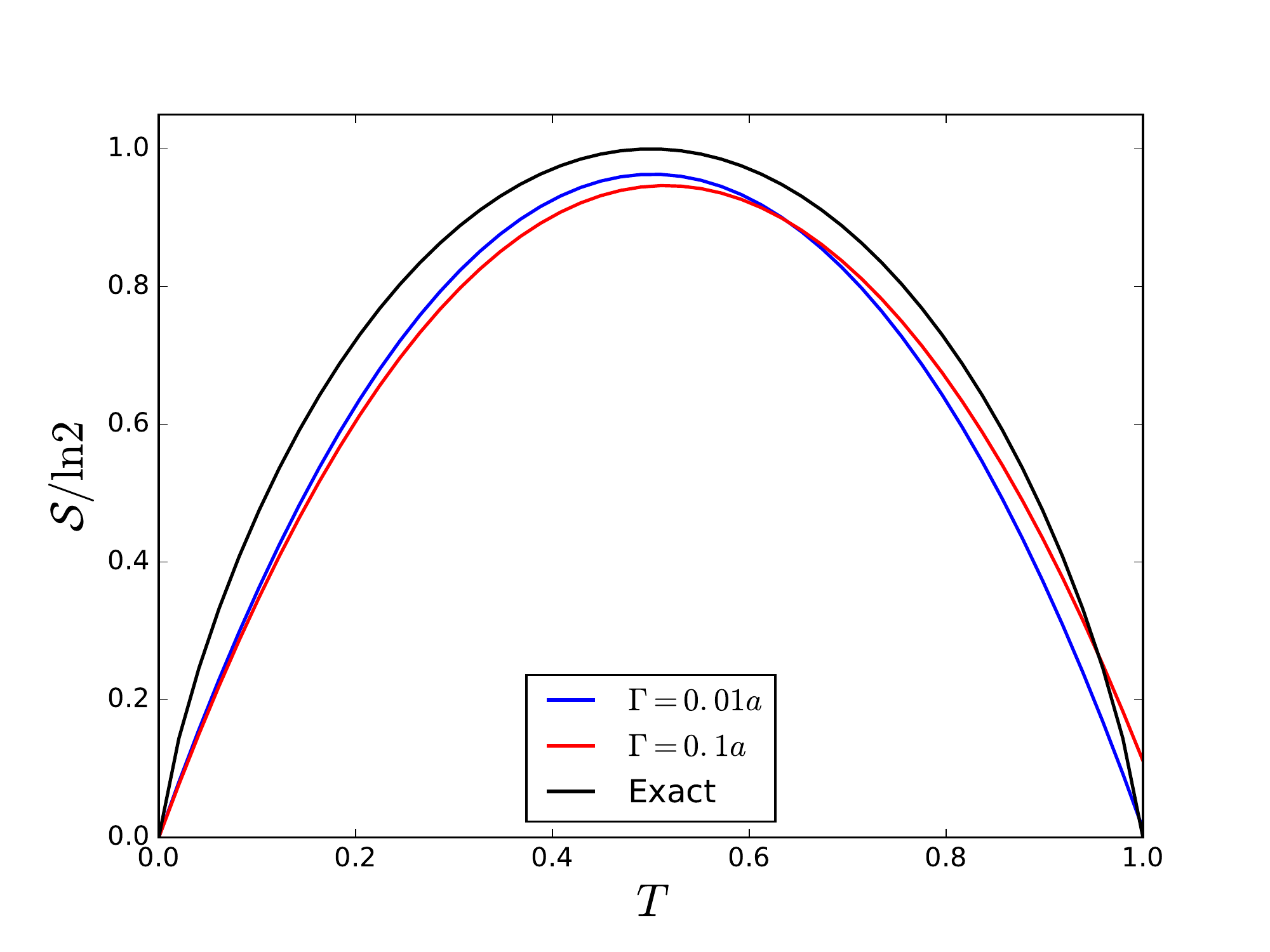}
  \caption{The entanglement entropy generated per period obtained from Eq.~(\ref{eq:entropy}). The exact result for the entanglement entropy reads $\mathcal{S}= (T-1)\ln(1-T)-T\ln T$. The maximum value $\mathcal{S}=\ln 2$ is obtained for $T=1/2$.}
  \label{fig:entanglement}
\end{figure}

\paragraph{Entanglement entropy.---} Finally, as an application of our scheme, we consider measuring the entanglement entropy generated by partitioning electrons on a QPC. Recently, it has been realized that the entanglement entropy between two electronic reservoirs connected by a QPC is closely linked to the FCS \cite{klich09,song11,song12,petrescu14,thomas15}. Specifically, the entanglement entropy generated per period can be approximated from the first four current cumulants as~\cite{song11,song12}
\begin{equation}
\mathcal{S}\simeq \alpha_2\langle\!\langle I^2\rangle\!\rangle+\alpha_4\langle\!\langle I^4\rangle\!\rangle,
\label{eq:entropy}
\end{equation}
where the coefficients $\alpha_m=2\sum_{k=m-1}^4 S_1(k,m-1)/(e^m k!k)$ are given by the unsigned Stirling numbers of the first kind $S_1(k,m)$.  Figure~\ref{fig:entanglement} shows that the entanglement entropy obtained from the cumulants in Fig.~\ref{fig:qpc}a is in good agreement with the exact result. This demonstrates that the entanglement entropy in a fermionic quantum many-body system may be within experimental reach.

\paragraph{Conclusions.---}
Electronic Mach-Zehnder interferometers can function as detectors of current fluctuations in mesoscopic conductors. Equation (\ref{eq:mgfmeasurement}) expresses the full counting statistics exclusively in terms of average currents measured at the outputs of the interferometer. Equation~(\ref{eq:lambdatau}) shows that the counting field can be controlled by varying the time delay between separate voltage signals. These findings make it possible to measure the current cumulants as well as the full distribution of current fluctuations as illustrated in Fig.~\ref{fig:qpc}. Our scheme is robust against moderate dephasing and finite temperature effects. As an application we have shown that our scheme enables measurements of the entanglement entropy in fermionic many-body systems. Extensions of our work may facilitate the detection of short-time observables such as the electronic waiting time distribution \cite{albert12,dasenbrook14,dasenbrook16} \revision{or include superconducting circuit elements~\cite{belzig01}}.

\paragraph{Acknowledgments.---}
We thank D.~Golubev, P.~P.~Hofer, G.~B.~Lesovik, and C.~Padurariu for useful comments. We thank P.~P.~Hofer for helping us produce Fig.~\ref{fig:setup}.  CF is affiliated with Centre for Quantum Engineering at Aalto University. DD acknowledges the hospitality of Aalto University. The work was supported by Swiss NSF and Academy of Finland.

\includepdf[pages={{},1}]{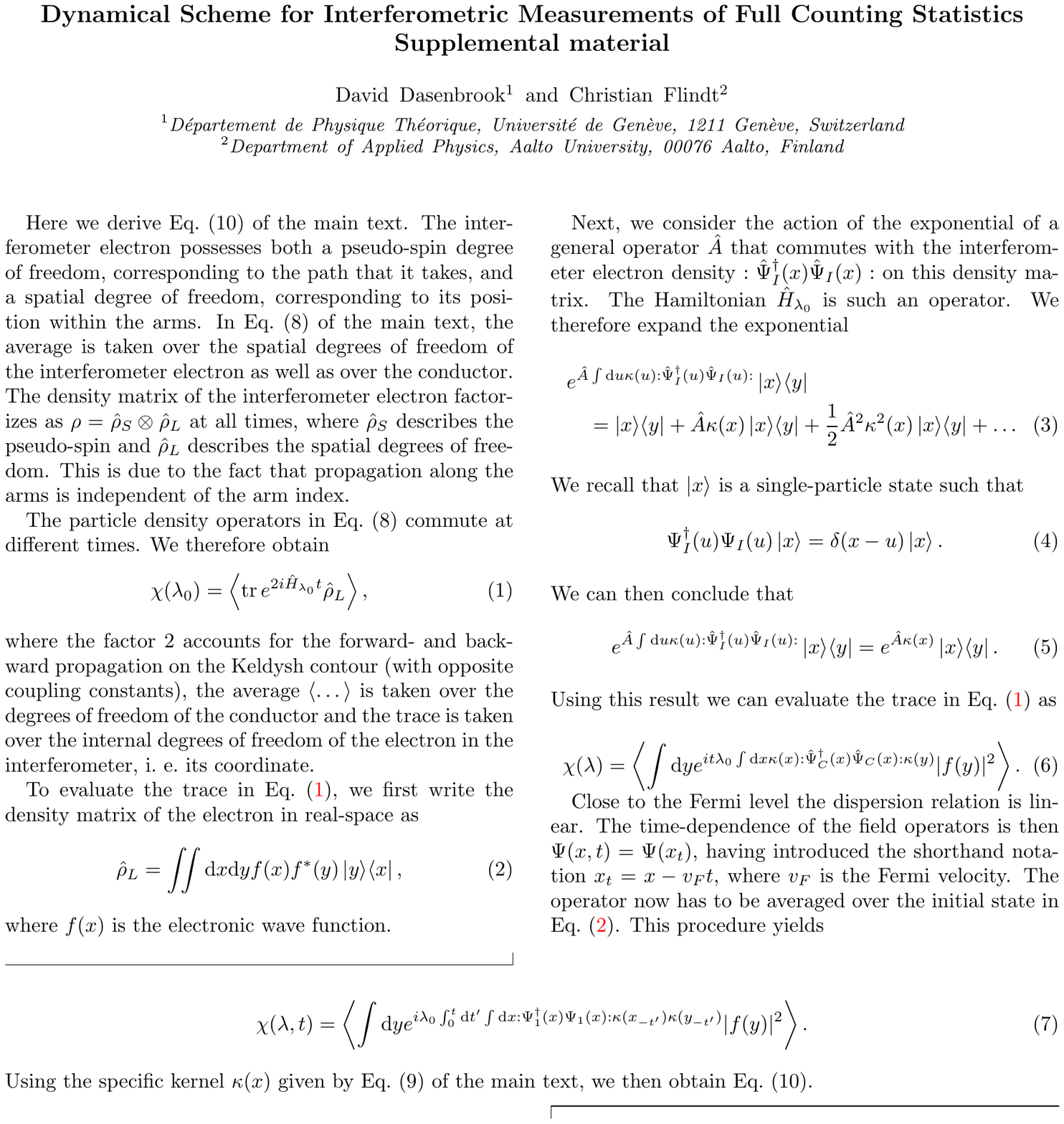}


\end{document}